\documentclass[aps,prl,reprint,amsmath,amssymb,superscriptaddress,showpacs]{revtex4-1}
\usepackage{bm}
\usepackage{graphicx}
\usepackage{subfigure}
\usepackage{color}
\usepackage{dcolumn}
\usepackage{kotex}
\usepackage[font=small,labelfont=bf,labelsep=space]{caption}
\usepackage{autobreak}
\begin{document}
	
\title{Orbital-selective confinement effect of Ru 4$d$ orbitals in SrRuO$_3$ ultrathin film}

\author{Soonmin Kang}
\affiliation{Center for Correlated Electron Systems, Institute for Basic Science (IBS), Seoul 08826, Korea}
\affiliation{Department of Physics and Astronomy, Seoul National University, Seoul 08826, Korea}

\author{Yi Tseng}
\affiliation{Swiss Light Source, Paul Scherrer Institut (PSI), CH-5232 Villigen, Switzerland}

\author{Beom Hyun Kim}
\affiliation{Korea Institute for Advanced Study (KIAS), Seoul 02455, Korea}

\author{Seokhwan Yun}
\affiliation{Center for Correlated Electron Systems, Institute for Basic Science (IBS), Seoul 08826, Korea}
\affiliation{Department of Physics and Astronomy, Seoul National University, Seoul 08826, Korea}

\author{Byungmin Sohn}
\affiliation{Center for Correlated Electron Systems, Institute for Basic Science (IBS), Seoul 08826, Korea}
\affiliation{Department of Physics and Astronomy, Seoul National University, Seoul 08826, Korea}

\author{Bongju Kim}
\affiliation{Center for Correlated Electron Systems, Institute for Basic Science (IBS), Seoul 08826, Korea}
\affiliation{Department of Physics and Astronomy, Seoul National University, Seoul 08826, Korea}

\author{Daniel McNally}
\affiliation{Swiss Light Source, Paul Scherrer Institut (PSI), CH-5232 Villigen, Switzerland}

\author{Eugenio Paris}
\affiliation{Swiss Light Source, Paul Scherrer Institut (PSI), CH-5232 Villigen, Switzerland}

\author{Choong H. Kim}
\affiliation{Center for Correlated Electron Systems, Institute for Basic Science (IBS), Seoul 08826, Korea}
\affiliation{Department of Physics and Astronomy, Seoul National University, Seoul 08826, Korea}

\author{Changyoung Kim}
\affiliation{Center for Correlated Electron Systems, Institute for Basic Science (IBS), Seoul 08826, Korea}
\affiliation{Department of Physics and Astronomy, Seoul National University, Seoul 08826, Korea}

\author{Tae Won Noh}
\affiliation{Center for Correlated Electron Systems, Institute for Basic Science (IBS), Seoul 08826, Korea}
\affiliation{Department of Physics and Astronomy, Seoul National University, Seoul 08826, Korea}

\author{Sumio Ishihara}
\affiliation{Department of Physics, Tohoku University, Sendai 980-8578, Japan}

\author{Thorsten Schmitt}
\email{thorsten.schmitt@psi.ch}
\affiliation{Swiss Light Source, Paul Scherrer Institut (PSI), CH-5232 Villigen, Switzerland}

\author{Je-Geun Park}
\email{jgpark10@snu.ac.kr}
\affiliation{Center for Correlated Electron Systems, Institute for Basic Science (IBS), Seoul 08826, Korea}
\affiliation{Department of Physics and Astronomy, Seoul National University, Seoul 08826, Korea}

\date{\today}

\begin{abstract}
The electronic structure of SrRuO$_3$ thin film with thickness from 50 to 1 unit cell (u.c.) is investigated via the resonant inelastic x-ray scattering (RIXS) technique at the O K-edge to unravel the intriguing interplay of orbital and charge degrees of freedom. We found that orbital-selective quantum confinement effect (QCE) induces the splitting of Ru $4d$ orbitals. At the same time, we observed a clear suppression of the electron-hole continuum across the metal-to-insulator transition (MIT) occurring at the 4 u.c. sample. From these two clear observations we conclude that QCE gives rise to a Mott insulating phase in ultrathin SrRuO$_{3}$ films. Our interpretation of the RIXS spectra is supported by the configuration interaction calculations of RuO$_{6}$ clusters.

\end{abstract}

\pacs{fill in later}

\maketitle

\section*{I. Introduction}
Orbital degree of freedom (DOF) is relatively less well understood among the four fundamental DOF of solid: charge, spin, lattice, and orbital. The role of the orbital DOF was originally recognized by the now famous Kugel-Khomskii model \cite{82-SPU}. It has since taken another decade before its full consequence was experimentally observed from numerous studies on so-called colossal magnetoresistance (CMR) manganites \cite{01-RMP}. The most direct effect of orbital DOF can be found in the so-called orbital ordering and the associated metal-insulator transition (MIT) with unique magnetic or structural transitions \cite{06-NMAT,18-PRX}. A more recent breakthrough in an understanding of orbital DOF is in the discovery of orbital-selective mechanism. It is now believed that several Ru and V oxides exhibit the phenomena that arise from the orbital-selective physics \cite{06-PRL-CSRO,18-PRX,06-PRB-V2O3,13-PRB-VO2}. One notable example is the orbital-selective Mott transition \cite{16-PNAS}.

The role of orbital DOF is typically enhanced for localised systems, i.e. with a larger $U$ term. So it becomes more prominent in 3$d$ transition metal oxides, which is why CMR manganite was the first system that was identified with orbital physics. Nevertheless, several Ru compounds were also reported to have rather unique features due to the orbital physics. Despite the progress of our understanding of orbital physics for Ru, an orbital-selective process still remains pretty much unexplored for Ru compounds although it was already suggested for the doping-dependent MIT of (Ca,Sr)$_2$RuO$_4$ \cite{06-PRL-CSRO,10-JPCM-MIT,09-PRL-CSRO}.

SrRuO$_3$ is a well-known member of the ruthenates family with a ferromagnetic phase below the Curie temperature of 165 K. Unlike other ferromagnetic materials, conductivity of bulk SrRuO$_3$ is high enough to make it a popular choice of electrode for various thin film samples with a stable perovskite structure \cite{04-APL}. At the same time, it is one of the rare itinerant ferromagnetic oxides, which has attracted significant interest in its own right \cite{96-JPCM, 08-JKPS}. For example, it has long been suspected that some kind of coupling between the lattice and spin degrees of freedom works for the ferromagnetic ground state. It was also found both theoretically and experimentally that RuO$_6$ octahedra of SrRuO$_3$ undergoes quite irregular ‘plastic’ distortion below the ferromagnetic transition temperature \cite{06-PRB-strain,13-JPCM-structure}. More recently, the unusual temperature dependence of the spin gap found by inelastic neutron scattering was attributed to a possible magnetic monopole in the $k$-space \cite{16-NCOMMS}. Interestingly, it is known too that the metallic phase of bulk SrRuO$_3$ is close to a transition between Fermi-liquid and non-Fermi-liquid states \cite{96-JPCM,98-PRB}. Another interesting point, more relevant to our work, is that SrRuO$_3$ thin films undergo MIT with decreasing thickness, whose origin is to date not well understood \cite{05-APL,09-PRB,09-PRL}. Thus, SrRuO$_3$ thin films can be a fertile ground for exploring some of the fundamental physics related to MIT and correlation physics with the orbital DOF.

In addition, first-principle LDA+U calculations found that the Ru orbitals of SrRuO$_3$ thin films exhibit rather unusual quantum confinement effects (QCE) when reducing thickness \cite{09-PRL}. As the thickness of film gets reduced, the proportion of RuO$_{6}$ octahedra exposed to the surface increases, which makes Ru $t_{2g}$ orbitals like $d_{xz}$ or $d_{yz}$ to prefer to form one-dimensional (1D) strips. As a result of the geometrical restriction, enhanced QCE was theoretically predicted to induce a distinctive change in the electronic structures for Ru 4$d$ orbitals. To be more specific, density of states (DOS) for a 2D square lattice with a tight-binding model has a van Hove singularity at the band center whereas DOS for a 1D line case has two separate singularities at the each edge of the band \cite{04-Cambridge}. For example, the 2D-type van Hove singularity of $d_{xy}$ DOS persists down to monolayer SrRuO$_3$. However, $d_{xz}$ and $d_{yz}$ orbitals in monolayer limit do not have electron hopping along the z-axis due to spacial confinement, which induces the 1D-type singularities of their DOS. This orbital-selective QCE was theoretically suggested to be the main driving force of the intriguing paramagnetic phase found for very thin SrRuO$_3$ samples \cite{09-PRL}. We also note that QCE was used to explain the Mott insulating phase of LaNiO$_3$/LaAlO$_3$ thin films \cite{11-PRB-LNOLAO}.

The purpose of this study was twofold. First, we investigated the proposed QCE by measuring the orbital-dependent charge transfer with the high-resolution RIXS studies as a function of thickness. Second, we studied how the charge dynamics changes across the MIT by examining low energy excitations across the critical thickness. Furthermore, we tried to find correlation between those two distinct characteristics of SrRuO$_3$ thin film. 

\section*{II. Experimental methods}
Epitaxial SrRuO$_3$ thin films were deposited on TiO$_2$-terminated SrTiO$_3$ (001) substrates by pulsed laser deposition (PLD) at 670$~^{\circ}\mathrm{C}$ with oxygen pressure of 100 mTorr. Ultraviolet light coming from the excimer laser with power of 2.1 $J/cm^2$ is applied to the target with a spot size of 2 $mm^2$. We optimized the growth condition by measuring the resistivity of our samples and thereby monitoring the quality in addition to the usual inspection of the reflection high-energy electron diffraction (RHEED) patterns. The RHEED pattern in time variation implies good surface quality, which shows a clear change in the growth mode from a layer-by-layer growth to a step flow growth as a function of time. On the other hand, the high residual resistivity ratio of 8.2 obtained for the samples testifies the high quality of our samples. In addition to the resistivity measurement, we verified the roughness of the samples in atomic force microscopy (AFM) images, another sign of the high quality of surface in thin films (Fig. 1).

\begin{figure}
	\centering
	\includegraphics[width=\columnwidth,clip]{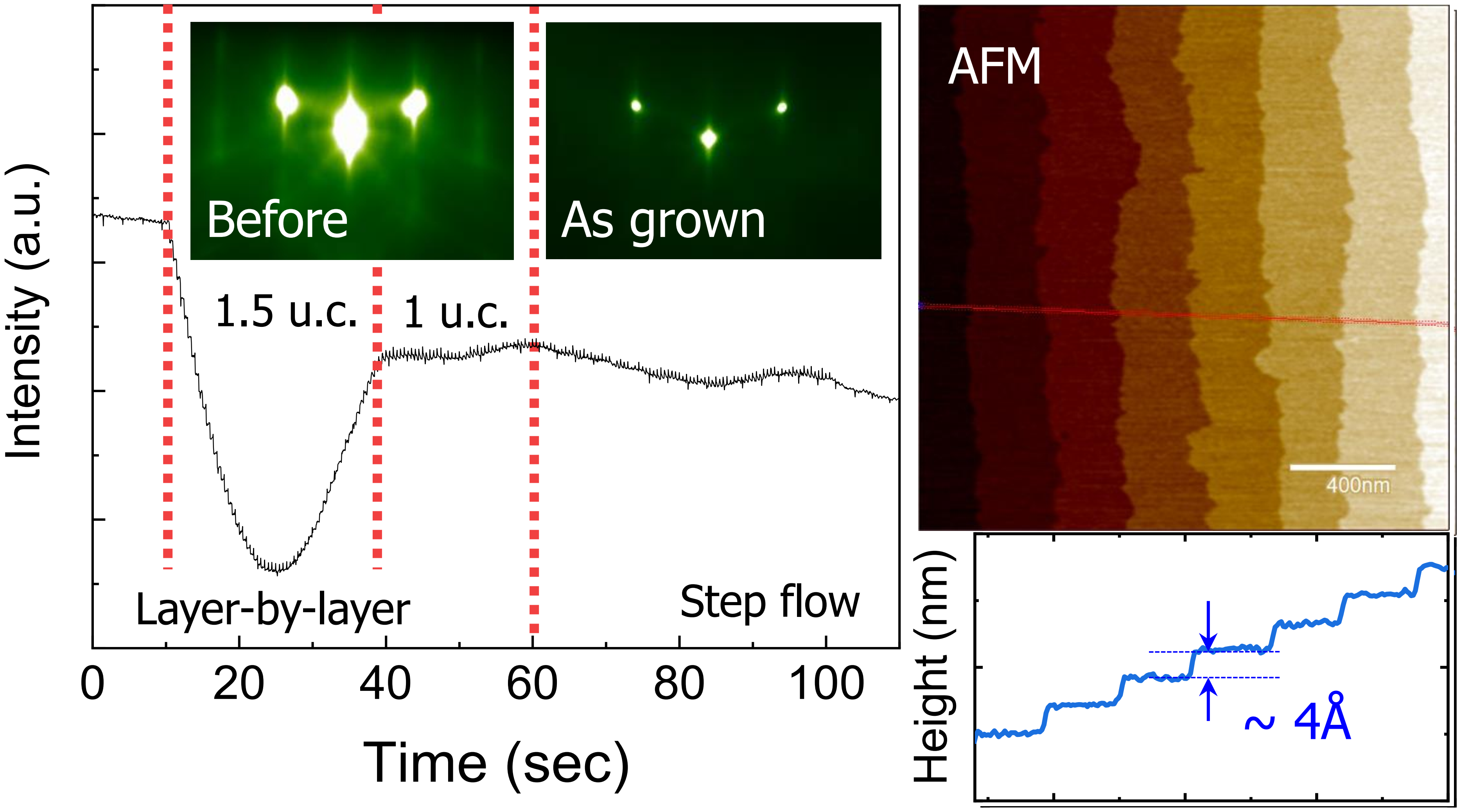}
	\caption{\label{Figure 1}(color online) In-situ RHEED pattern and topography image with AFM. The sample growth starts from 10 seconds. Growth mode change that occurs at 40 and 60 seconds shows the good surface quality of thin films. Inset figures show RHEED patterns before and after the sample growth. AFM image indicates the clean surface and the apparent steps with the height of 4 \AA, which is the size of 1 u.c. for SrRuO$_3$.}
\end{figure}

We carried out O K-edge resonant inelastic x-ray scattering (RIXS) at the ADRESS beamline of Swiss Light Source \cite{10-JSR, 06-RSI}. RIXS is a powerful tool to study the charge dynamics related to orbital physics as one can tune the energy to a specific absorption resonance of elements. Energy of ruthenium L-edge ($\sim 3$ keV), however, just happens to be situated in between soft and hard x-ray regimes. Because of this technical reason, it is not easy to get enough photon flux and energy resolution at the Ru L-edge, which makes it experimentally challenging to do RIXS at the Ru L-edge. Instead, we carried out our experiment at the oxygen K-edge to study the charge dynamics of the Ru $4d$ orbitals while varying the thickness of thin film samples.

The proper energy of the incident beam was chosen through x-ray absorption spectroscopy (XAS) with different thickness from 1 to 33 u.c. as shown in Fig. 2. The first peak at around 529.8 eV gets weaker as the thickness of the samples becomes reduced. From the fact that the relative intensity changes for different samples and also based on the previous XAS studies in SrTiO$_3$ \cite{13-PRL_STO, 07-PRB_STO}, we conclude that peaks at above 530 eV are due to absorptions from the substrates. Therefore we chose 529.8 eV as an incident energy for our RIXS experiment with high statistics, which is slightly lower than the pure O K-edge. The energy difference between pure O K-edge and absorption from our sample comes from the hybridization energy. We verified the energy resolution to be less than 70 meV by checking the full width at half maximum of the elastic line from diffuse scattering at a carbon tape reference.

\begin{figure}
	\centering
	\includegraphics[width=\columnwidth,clip]{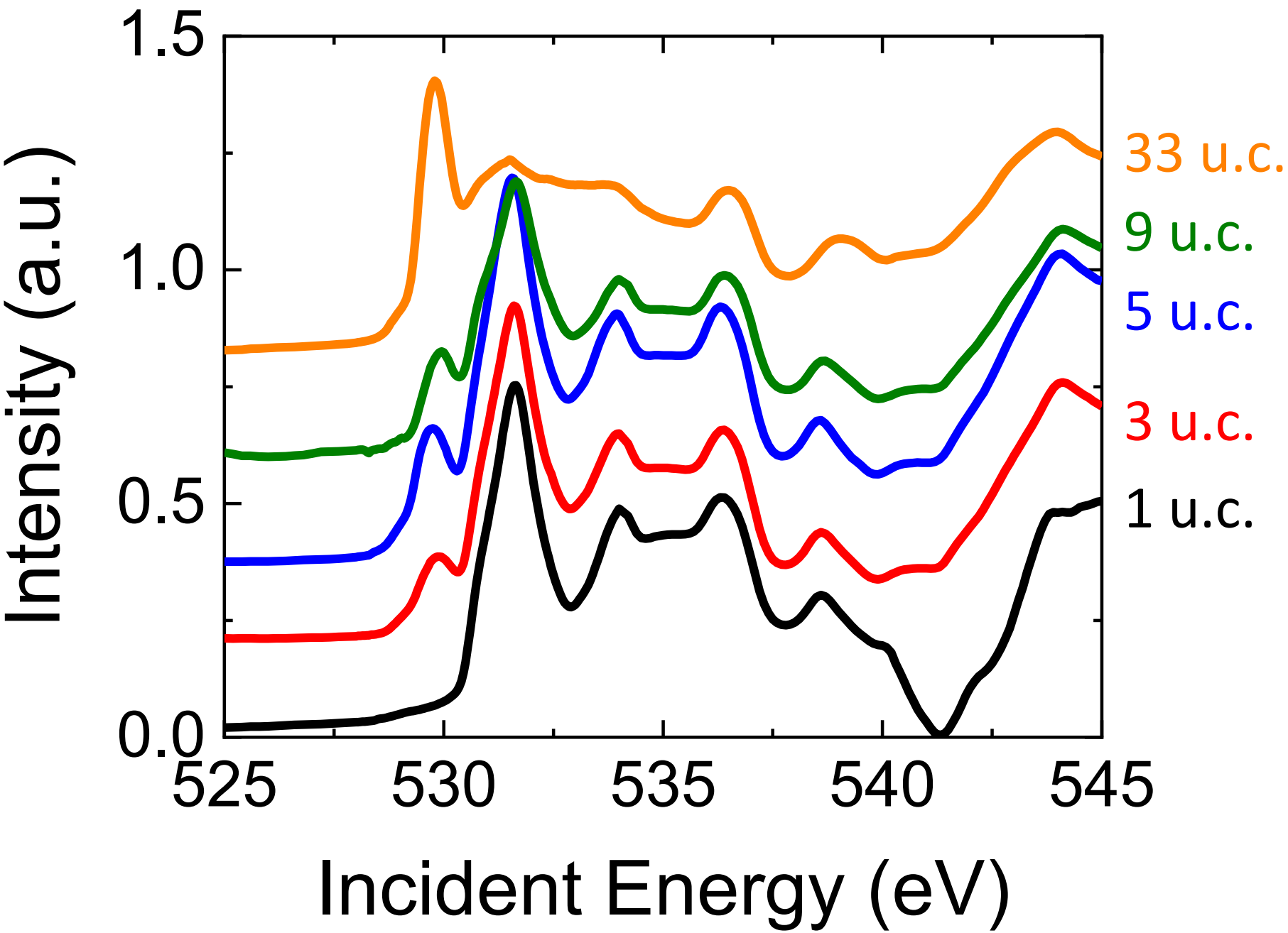}
	\caption{\label{Figure 2}(color online) XAS results as a function of the thickness of the sample. The energy of 529.8 eV was used for our RIXS experiments because other peaks mainly originate from the SrTiO$_3$ substrate. }
\end{figure}

All our samples were aligned with a grazing angle ($\theta = 15~^{\circ}$) to increase the scattering cross section especially for ultrathin samples. The scattering angle from incident beam to detector was fixed to 130$~^{\circ}$, with the corresponding momentum transfer of $\mathbf{q}_{\|}$ = 0.28 [$2\pi / a$]. We employed two different polarizations for our experiments: $\sigma$ polarization is parallel to the sample plane and $\pi$ polarization is nearly perpendicular to the plane. Thus, the former is more sensitive to $p_x$($p_y$) orbital while the latter is so to $p_z$ orbital due to the incident angle. All experiments were performed at 20 K.

\begin{figure}
	\centering
	\includegraphics[width=\columnwidth,clip]{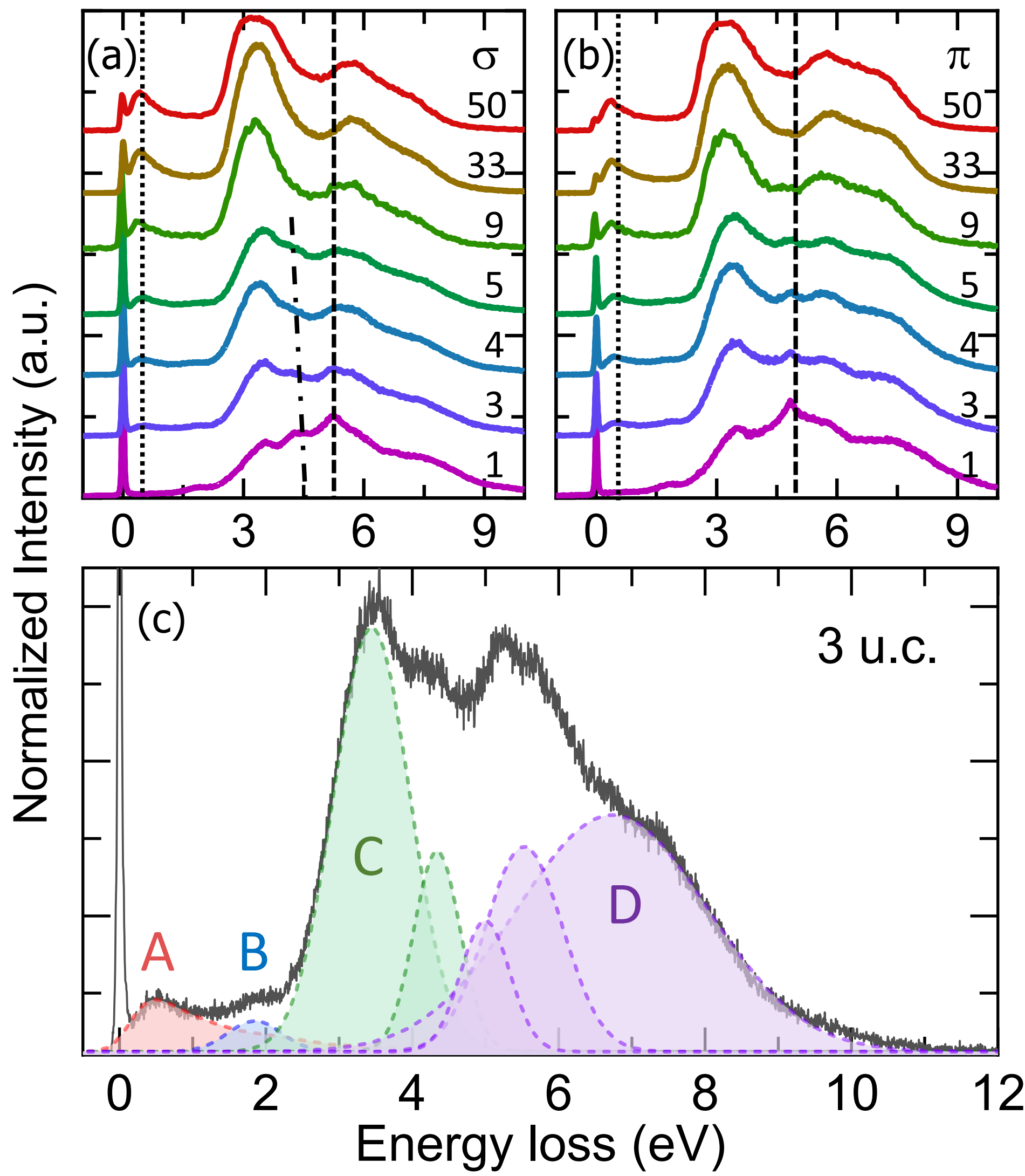}
	\caption{\label{Figure 3}(color online) (a, b) RIXS spectra at the O K-edge with $\sigma$ and $\pi$ polarizations for SrRuO$_3$ thin films. Upper figures show the overall features of RIXS spectra depending on the thickness of the samples and the polarization of incident beam. As the thickness decreases, the peak at the low energy side (dot line) becomes weaker while the peak at 5 eV(dashed line) gets stronger for both polarizations. In addition to the 5 eV peak, the peak around 4.5 eV(dash-dot line) also appears for the $\sigma$ polarization. Note that this 4.5 eV peak becomes stronger below 5 u.c. sample and shifts towards higher energy as decreasing the thickness. (c) The lower graph shows the whole spectrum for the 1 u.c. sample with the $\sigma$ polarization. Altogether seven Gaussian fitting functions are needed to fit the spectra based on the CI and DFT calculations. Different types of peak are marked by different alphabet in the lower graph.}
\end{figure}

Fig. 3 shows RIXS results for all seven samples with different thickness. To explain the RIXS spectra, we divided the spectra into two groups depending on the characteristic energy of the peaks and their apparent relevance to our two main questions: QCE and MIT, respectively. For example, in the high energy side ranging from 2 to 10 eV there are several strong peaks marked as C and D, respectively. These peaks are due to the charge transfer from O $2p$ to Ru $4d$ orbitals and so reflect the expected change in the Ru $4d$ orbitals. On the other hand, there are two relatively weaker peaks below 2 eV with strong thickness dependence. These low-energy excitations can be interpreted as arising from d-d excitations or coherent peaks connected to quasiparticle states that are closely related to the metallic phase of SrRuO$_3$. In the remaining part of the paper, we would like to focus on the charge transfers to explain QCE first and then move on to the low energy part for MIT. 

\begin{figure}
	\centering
	\includegraphics[width=\columnwidth,clip]{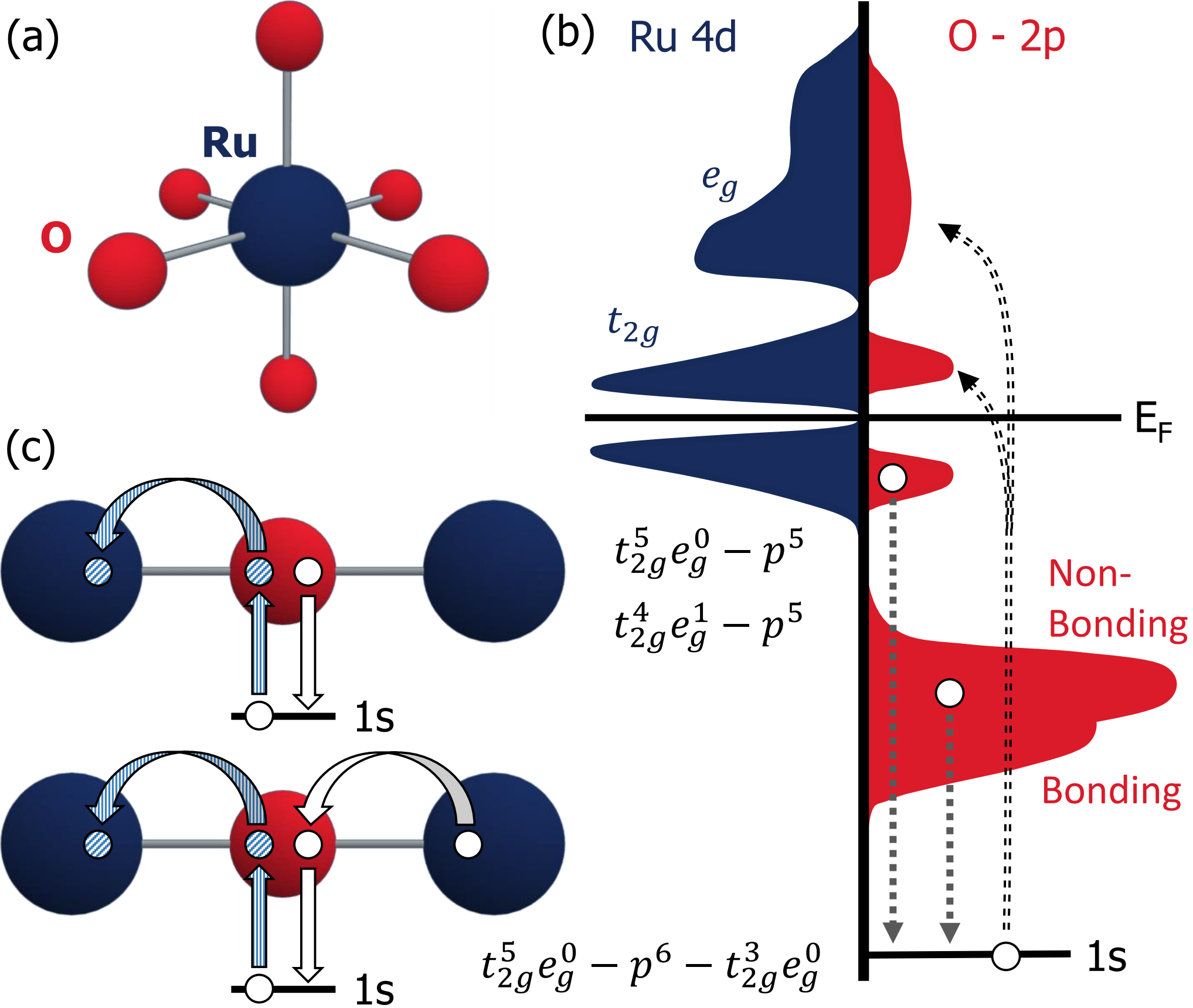}
	\caption{\label{Figure 4}(color online) (a) RuO$_6$ cluster used in the CI calculation. (b, c) Schematic view of charge transfer and d-d excitations. Oxygen $1s$ electrons are excited to vacant $2p$ levels which are hybridized with Ru $t_{2g}$ orbitals. The energy losses should be different depending on orbitals from which the relaxation occurs. Top figure in (c) shows $t_{2g}-e_{g}$ excitations and the bottom one indicates the charge transfer from O $2p$ to Ru $4d$ levels.}
\end{figure}

\section*{III. Results and Discussion}
\subsection*{A. Configuration interaction calculation of cluster models}
In order to explain the charge transfer peaks and d-d excitations in detail, we performed the configuration interaction (CI) calculations using two cluster models of RuO$_6$ and Ru-O-Ru (see Fig. 4a) to find that each of the calculations with different clusters shows distinct features of SrRuO$_3$. We note that our model calculation suits for $t_{2g}$ orbitals of more localized character. For instance, this calculation with the RuO$_6$ cluster model has advantage in explaining the charge transfer between O $2p$ and Ru $d$ orbitals because the cluster consists of six oxygen atoms. On the other hand, the calculation with the Ru-O-Ru cluster gives a better description of intersite d-d excitations. These calculations can also reflect the QCE by the extra control of adjusting the amount of Ru $d$ splitting. For example, we can set Ru $d$ orbitals to split into $\varepsilon_{xy} = 2/3\Delta_{t_{2g}}$, $\varepsilon_{xz} = \varepsilon_{yz} = -1/3\Delta_{t_{2g}}$, $\varepsilon_{z^{2}} = 10Dq - 1/2\Delta_{e_{g}}$, and $\varepsilon_{x^{2}-y^{2}} = 10Dq + 1/2\Delta_{e_{g}}$. It is to be noted that we used an unusually large energy splitting between $d_{xy}$ and $d_{xz}$($d_{yz}$) orbitals ($\Delta_{t_{2g}}$=0.8 eV) from the results of first-principle calculation in ref. \cite{09-PRL}, which is the energy difference between the 2D-type singularity of $d_{xy}$ and the 1D-type singularity of $d_{xz}$($d_{yz}$). 

\begin{table*}[!t]
	\centering
	\renewcommand{\arraystretch}{1.4}
	\caption{Physical parameters used for the cluster calculations in units of eV.}
	\begin{tabular*}{\linewidth}{@{\extracolsep{\fill}}cccccccccc}
		\hline\hline
		$10Dq$ & $\Delta_{t_{2g}}$ & $\Delta_{e_{g}}$ & $\lambda$ & $U$ & $J_{H}$ & $\Delta$ & $\Delta_{p}$ & $V_{pd\sigma}$ & $V_{pd\pi}$ \\
		\hline
		2.1 & 0.8 & 0.4 & 0.1 & 2.0 & 0.3 & 3.3 & 1.6 & -1.0 & 0.46 \\ 
		\hline\hline
	\end{tabular*}%
	\label{table:1}
\end{table*}

We also take into account both the spin-orbit coupling ($\lambda$) and the Kanamori-type Coulomb interaction ($U$ and $J_{H}$) among $d$ orbitals \cite{15-PRB-BHKim}. The energy levels of oxygen $p$ orbitals in the valence band can depend on whether they are hybridized with Ru $d$ orbitals or not \cite{04-PRB-PES, 05-PRB-PES, 05-PRB-U/W}. For example, O $p$ orbitals are assumed in our calculations to be non-interacting and their energy levels are given as $e_{p}$ for non-bonding and $e_{p} - \Delta_{p}$ for bonding $p$ orbitals, respectively. $e_{p}$ is determined as $e_{p} = 4U - 7J_{H} - \Delta$, where $\Delta$ is the charge transfer energy in the cubic symmetry defined as the energy difference between lowest $d^{5}\underline{L}$ and $d^{4}$ states. The hopping integrals between $p$ and $d$ orbitals are parameterized with $V_{pd\pi}$ for $t_{2g}$ and $V_{pd\sigma}$ for $e_{g}$ orbitals according to the Slater-Koster theory \cite{54-PR}. We used the parameters shown in Table \ref{table:1} in order to fit the experimental RIXS spectrum.

For more details of our calculations, let $| \Psi_g \rangle$ and $E_g$ be the ground state and its energy, respectively. In the dipole and fast collision approximation, the oxygen K-edge RIXS intensity at zero momentum is given as
\begin{equation}
	\begin{split}
	I\sim -\frac{1}{\pi}\textrm{Im}\langle \Psi_g | \hat{R}(\boldsymbol{\epsilon},{\boldsymbol{\epsilon}}')\frac{1}{\omega-H+E_g+i\delta}\hat{R}(\boldsymbol{\epsilon},{\boldsymbol{\epsilon}}') | \Psi_g\rangle.
	\end{split}
\end{equation}
And $\hat{R}(\boldsymbol{\epsilon},{\boldsymbol{\epsilon}}')$ is the RIXS scattering operator given as
\begin{equation}
	\begin{split}
	\hat{R}(\boldsymbol{\epsilon},{\boldsymbol{\epsilon}}')=\frac{1}{3}\sum_{imm'\sigma}\epsilon_m \epsilon_{m'}' c_{im'\sigma} c_{im'\sigma}^\dagger,
	\end{split}
\end{equation}
where $c_{im'\sigma}^\dagger$ is the creation operator of oxygen p electron with $m=(x,y,z)$ orbital and $\sigma$ spin at an $i$-th site, and $\boldsymbol{\epsilon}$ and $\boldsymbol{\epsilon}'$ are the polarizations of incident and outgoing x-rays, respectively\cite{15-PRB-BHKim}. $\delta$ is the Lorentz broadening and we set $\delta=0.2$ eV for our calculations.

For $p$ orbital states, they can be expressed with a linear combination of bonding and non-bonding states like
\begin{equation}
	\begin{split}
	c_{im\sigma}^\dagger = \sum_{\alpha}(U_{\alpha,im}^{B})^{*}c_{\alpha\sigma}^\dagger + \sum_{\mu}(U_{\mu,im}^{N})^{*}c_{\mu\sigma}^\dagger,
	\end{split}
\end{equation}
where $U_{\alpha,im}^{B}$ and $U_{\mu,im}^{N}$ are the coefficients of $m$ orbital at the $i$-th site for bonding and non-bonding states $\alpha$ and $\mu$, respectively. Because non-bonding $p$ orbitals are fully occupied in the ground state, only annihilation operation is allowed. We can then get the scattering operator associated with non-bonding orbitals as following
\begin{equation}
	\begin{split}
	\hat{R}^N(\boldsymbol{\epsilon},\boldsymbol{\epsilon}') = \sum_{\alpha\mu\sigma}R_{\alpha\mu}^{N}(\boldsymbol{\epsilon},\boldsymbol{\epsilon}')c_{\mu\sigma}c_{\alpha\sigma}^{\dagger},
	\end{split}
\end{equation}
where $R_{\alpha\mu}^{N}=\!\frac{1}{3}\sum_{imm'} U_{\mu,im'}^{N}(U_{\alpha,im}^B)^* \epsilon_{m'}' \epsilon_{m}$. The RIXS intensity attributed to non-bonding $p$ orbitals is given as
\begin{multline}
	I^{N}=-\frac{1}{\pi}\textrm{Im}\sum_{\alpha\alpha'\mu} R_{\alpha'\mu}^{N}(\boldsymbol{\epsilon},\boldsymbol{\epsilon}')^* R_{\alpha\mu}^{N}(\boldsymbol{\epsilon},\boldsymbol{\epsilon}') \\
	\times \langle \Psi_g | c_{\alpha'\sigma} \frac{1}{\omega-H+E_g+e_p+i\delta} c_{\alpha'\sigma}^{\dagger} | \Psi_g\rangle.
\end{multline}
The RIXS intensity attributed to the bonding $p$ orbitals can then be calculated using the following relation
\begin{equation}
	\begin{split}
	I^{B} = -\frac{1}{\pi}\textrm{Im}\langle \Psi_g | \hat{R}^{B}(\boldsymbol{\epsilon},{\boldsymbol{\epsilon}}')\frac{1}{\omega-H+E_g+i\delta}\hat{R}^{B}(\boldsymbol{\epsilon},{\boldsymbol{\epsilon}}') | \Psi_g\rangle ,
	\end{split}
\end{equation}
where
\begin{equation}
	\begin{split}
	\hat{R}^{B}(\boldsymbol{\epsilon},{\boldsymbol{\epsilon}}') = \frac{1}{3} \sum_{\alpha \beta \sigma i m m'} U_{\beta,im'}^{B} (U_{\alpha,im}^{B})^{*} \epsilon_{m'}' \epsilon_m c_{\beta\sigma} c_{\alpha\sigma}^{\dagger}.
	\end{split}
\end{equation}

In case of the CI calculation of a Ru-O-Ru cluster, mainly explaining the low energy excitations, we directly used Eqs. (1) and (2) instead of considering bonding and non-bonding states. In addition, we restricted the Hilbert space with the following assumption that the oxygen atom between two Ru atoms has three possible states of $p^4$, $p^5$, and $p^6$ electron configurations. The result of this calculation is shown in Fig. 5.

\begin{figure}
	\centering
	\includegraphics[width=\columnwidth,clip]{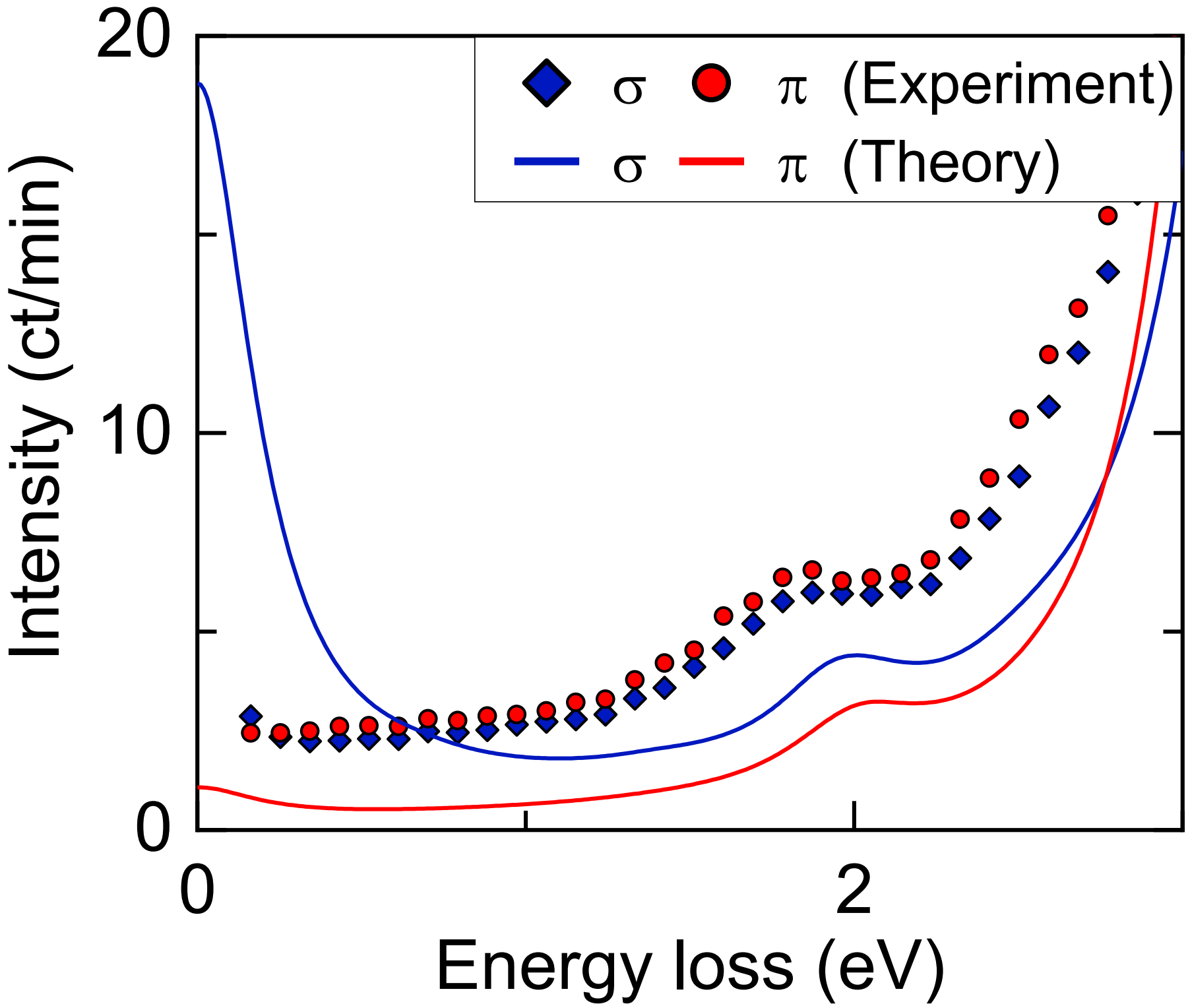}
	\caption{\label{Figure 5}(color online) Low energy RIXS spectra of 1 u.c. SrRuO$_3$ with the CI calculation of a Ru-O-Ru cluster. The symbols represent the experimental results while the lines show the theoretical results with different color used for the different polarization of the incident beam. The peak at 2 eV shows intersite d-d excitations.}
\end{figure}

The peaks in the O K-edge RIXS spectrum can also be categorized according to Ru $4d$ orbitals that participate in the RIXS process as shown in Fig. 4. Electrons in the core oxygen levels are excited to vacant O $2p$ levels that are hybridized with Ru $4d$ orbitals as seen in the O K-edge RIXS and subsequent relaxation occurs from the occupied $2p$ states. We can, in principle, determine the origin of each peak by examining the energy of the emitted photons. For example, if the electrons are relaxed from $2p$ level hybridized with $t_{2g}$ levels that are located right below the Fermi level, the process can be considered as d-d excitations. In the case of charge transfers between $2p$ and $4d$ orbitals, however, the relaxation starts from $2p$ states not participating in the hybridization.

\subsection*{B. Quantum confinement effects}
According to our CI calculations, the charge transfers correspond to the peaks C and D as observed from 2 to 10 eV. Peak C, for instance, represents the charge transfer between non-bonding O $2p$ states and Ru $t_{2g}$ orbitals while peak D mainly originates from bonding O $2p$ states and Ru $e_g$ orbitals. As shown in top graphs of Fig. 6, both peaks C and D undergo a considerable change depending on the thickness of the sample and the polarization of incident beam. The remarkable change of peak C is clearly seen around 4.4 eV. It is notable that this variation only occurs for the $\sigma$ polarization. Meanwhile, an additional peak emerges around 5 eV that is most likely due to the charge transfer between O $2p$ and Ru $e_{g}$ levels in both polarization channels, but the position of the peak is slightly different depending on the polarization (see Fig. 6).

\begin{figure}
	\centering
	\includegraphics[width=\columnwidth,clip]{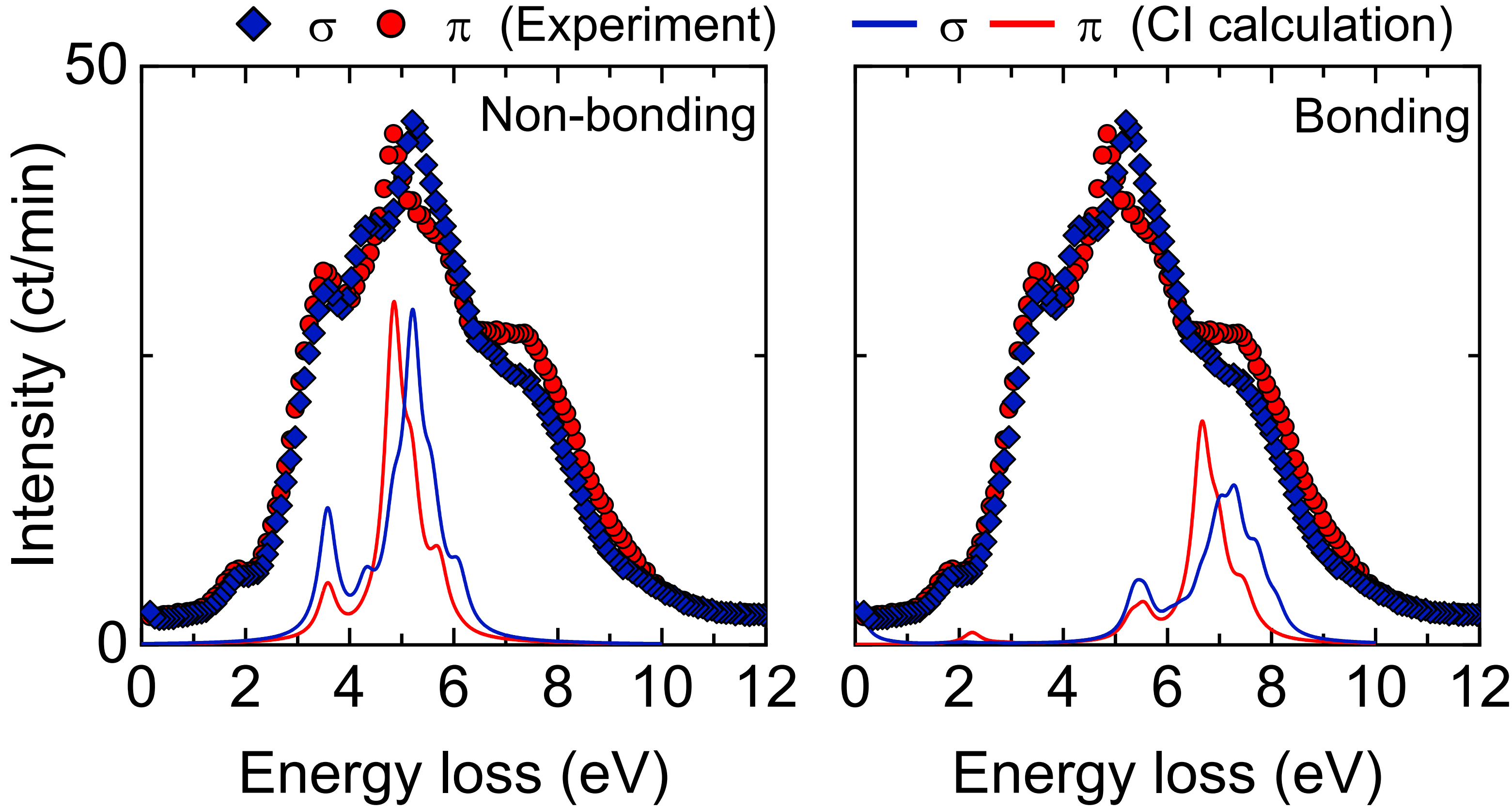}
	\caption{\label{Figure 6}(color online) RIXS spectra with the results of CI calculation for monolayer SrRuO$_3$. The symbols represent the experimental results while the lines show the theoretical results with different color used for the different polarization of the incident beam. (Left) Calculation results with non-bonding $p$ orbitals: (right) calculation results with bonding states.  }
\end{figure}

The splitting of both peaks shown in Figs. 3 and 6 can be taken as the evidence of QCE, which is more pronounced for the thinner samples. The splitting of peaks around 4 and 5 eV reflects the energy splitting of Ru $t_{2g}$ and $e_{g}$, respectively. Of interest, the QCE in monolayer SrRuO$_3$ modifies the electronic structure, which subsequently induces the separate orbital energy levels depending on the geometrical characteristics of each orbital. We comment that the energy difference between each singularity of the 2D-type band for $d_{xy}$ and the 1D-type band of $d_{xz}$($d_{yz}$) corresponds quite well to the amount of peak splitting in peak C \cite{09-PRL}. It should also be noted that 0.8 eV of $t_{2g}$ energy splitting cannot be obtained in the cases of the usual Jahn-Teller distortion: which is typically about 0.1 eV for $t_{2g}$ of ruthenates \cite{13-frontier}. 

A further interesting point is the polarization dependence of the peaks. In our explanation, the QCE pushes the energy levels of $d_{xz}$($d_{yz}$) or $d_{z^2}$ down so that the energy of charge transfer related to those orbitals gets shifted towards lower energy. On the other hand, orbitals such as $d_{xy}$ or $d_{x^2-y^2}$ move in the opposite direction. In the case of the charge transfer between $d_{xz}$($d_{yz}$) and $p$ orbitals, the same amount of energy shift compensates for the hopping integral $V_{pd\pi}$. Thus the additional peak at 4.4 eV appears only with the orbitals parallel to the surface of the samples and the one around 5 eV emerges at different energy depending on the polarization of the incident beam. Because each polarization excites different O $p$ orbitals, we believe the `orbital-selective' characteristic of the QCE results in the observed polarization dependence.

\subsection*{C. Metal-insulator transition}
While the peaks related to the charge transfer seem to support our scenario of the QCE process in SrRuO$_{3}$ films, the ones in the low energy range produce the clearest evidence of MIT. For instance, with reducing the thickness peak A is suppressed rapidly but peak B gets enhanced simultaneously below the thickness of 5 u.c. This opposite trend of these two peaks A and B can be easily understood in terms of MIT as seen in the resistivity data shown in Fig. 7. We note that the critical thickness can depend on the growth conditions according to our fabrication of several SrRuO$_3$ films used for this work.

According to our CI calculations, peak B can be ascribed to d-d excitations between intersite $t_{2g}$ orbitals (Fig. 4c). Electrons are excited to O $2p$ levels that hybridize with Ru $t_{2g}$ levels in the valence band and afterwards relaxation occurs from the $t_{2g}$ levels in the conduction band. Although the process can, in principle, involves oxygen $p$ levels, it is intrinsically the excitations between two separate $t_{2g}$ bands in the valence and conduction bands. 

Meanwhile, the origin of peak A can be found by calculating the joint density of states (JDOS) from first-principle calculations with density functional theory. Joint density of states (JDOS) represents the probability of allowed interband transitions including absorption or energy-loss functions \cite{76-JDOS-JPC,16-JDOS}. We calculated JDOS by considering the energy levels in the valence and conduction bands. In our calculation, JDOS is given as

\begin{equation}
	\begin{split}
	J(\boldsymbol{q})=\sum_{\vec{k}}\delta(|\varepsilon_f(\boldsymbol{k})-\varepsilon_i(\boldsymbol{k}-\boldsymbol{q})|).
	\end{split}
\end{equation}

According to our experimental geometry with a grazing angle, we choose the interband transition with the fixed momentum transfer of $\mathbf{q}_{\|}$ = 0.28 [$2\pi / a$] and computed the DOS of the energy difference between two levels, which represent the theoretical spectrum of electron-hole excitations. By comparing our calculation results with the experimental data as shown in Fig. 7, the calculated JDOS for the electron-hole continuum is in good agreement with the lowest peak seen in bulk SrRuO$_{3}$. It means that peak A corresponds to itinerant quasi-particle excitations while peak B does to excitations between lower and upper Hubbard bands. In this sense, the spectral weight transfer from peak A to peak B is in good agreement with the MIT in SrRuO$_3$ thin films. We comment that the transfer of spectral weight from peak A to peak B is also consistent with MIT as seen in the resistivity data.

\begin{figure}
	\centering
	\includegraphics[width=\columnwidth,clip]{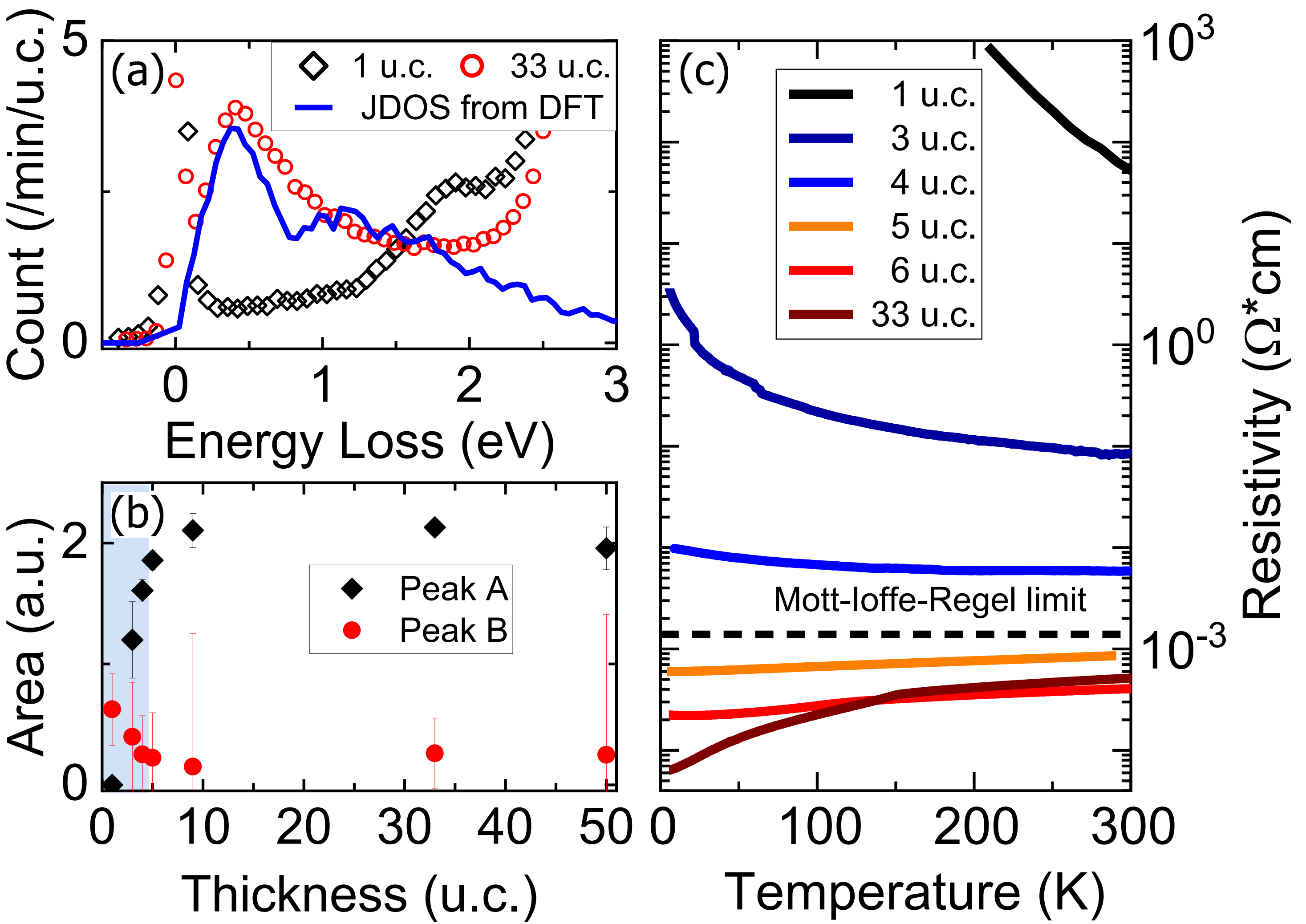}
	\caption{\label{Figure 7}(color online) (a, b) Low energy excitations are compared to the JDOS from DFT calculation. We clearly observe the electron-hole continuum in the thick sample, which arise from its metallic phase. The intensity of peak A sharply decreases below 5 u.c. and it completely disappears for the monolayer SrRuO$_3$. (c) Electical resistivity of SrRuO$_3$ thin films with different thickness. The resistivity increases progressively with reducing the thickness and crosses the theoretical Mott-Ioffe-Regel limit between 4 and 5 u.c. It is notable that the critical thickness from RIXS and resistivity coincides with one another. }
\end{figure}

Another interesting point is the connection between QCE and MIT, whose experimental evidence can be readily found in the very thin SrRuO$_{3}$ sample. In particular, a new peak is seen to be separated from the $d_{xy}$ level below 5 u.c. and moves towards higher energy as shown in Fig. 6. This means that QCE gets enhanced in thinner SrRuO$_3$ samples. With QCE splitting the Ru $4d$ bands, MIT in SrRuO$_{3}$ resembles that of Ca$_{2}$RuO$_{4}$, which is a classic example of an orbital-selective Mott insulator \cite{17-NCOMM-MottInsulator-Ca2RuO4}. For our thinnest sample of 1 u.c. SrRuO$_3$, QCE seems to split the otherwise degenerate $t_{2g}$ orbitals leading to a Mott-type insulating state. Therefore, we can maintain that a new way of realizing a Mott-type insulating phase is found in the ultrathin SrRuO$_3$ sample with thickness being a control parameter, which is different from the bulk sample.

\section*{IV. Conclusion}
To conclude, the good agreement between the theoretical calculation and the experimental observation of charge-transfer peak splitting in the RIXS spectra suggests the orbital-selective QCE in ultrathin SrRuO$_3$ film. We also found that the suppression of the low-energy excitations that arise from electron-hole continuum across the metal-insulator transition. Finally, our studies provide the clear experimental evidence that QCE leads to a Mott insulating phase in ultrathin SrRuO$_{3}$.

\section*{V. Acknowledgements}
We would like to acknowledge Daniel Khomskii and Bumjoon Kim for helpful discussion. The work at IBS CCES is supported by Institute of Basic Science (IBS) in Korea (Grant No. IBS-R009-G1, No. IBS-R009-G2, and IBS-R009-D1). The work at PSI is supported by the Swiss National Science Foundation through the NCCR MARVEL and the Sinergia network Mott Physics Beyond the Heisenberg Model (MPBH). We also thank Korea Institute for Advanced Study for providing computing resources (KIAS Center for Advanced Computation Linux Cluster System) for this work.

\end{document}